\documentclass[pre,twocolumn,showpacs]{revtex4}
\usepackage{epsf,colordvi,color,graphicx,epsfig}
\usepackage[latin1]{inputenc}
\usepackage{float}
\usepackage{amsfonts,amssymb,amsmath,wasysym}
\newcommand{\imgwidth}{0.85}

\begin{document}

\title{Response of Boolean networks to perturbations}
\author{C. Fretter and B. Drossel}
\affiliation{ Institut für Festkörperphysik, TU Darmstadt, Hochschulstraße 6, 64289 Darmstadt, Germany}

\begin{abstract}
We evaluate the probability that a Boolean network returns to an
attractor after perturbing $h$ nodes. We find that the return
probability as function of $h$ can display a variety of different
behaviours, which yields insights into the state-space structure. In
addition to performing computer simulations, we derive analytical
results for several types of Boolean networks, in particular for
Random Boolean Networks. We also apply our method to networks that have
been evolved for robustness to small perturbations, and to a biological example.
\end{abstract}
\maketitle

\section{Introduction}

Boolean networks (BN) are used to model the dynamics of a wide variety
of complex systems, ranging from neural networks \cite{rosen-zvi} and
social systems \cite{moreira04} to gene regulation
networks \cite{lagomarsino}, sometimes combined with evolutionary
processes \cite{bassler04}.  BN are composed of interacting
nodes with binary states, typically $0$ and $1$, coupled among each
other.  The state of each node evolves according to a function of the
states observed in a certain neighbourhood, similar to what is done
when using cellular automata \cite{herrmann88}, but in contrast to
cellular automata, BN have no regular lattice structure, and not all
nodes are assigned the same update function.

Usually, BN models are studied using deterministic dynamical
rules. After a transient time, these networks reach attractors, which
are a sequence of periodically repeated states. The number and length
of such attractors is an important property investigated in BNs. Of
equal importance are the sizes of the basins of attraction of these
attractors, which are the sets of states leading to the attractors.
Some networks may contain periodic sequences in state space that are
not ``attractors'' in the strict sense, because there are no states
outside this sequence that are attracted to it. However, in this paper
we do not make this distinction, and we call all sequence periodically
repeated sequences of states ``attractor''. 

Real networks are often influenced by noise, since molecule
concentrations may be small (e.g., in biological systems
\cite{arkin97}) or behaviour may be unpredictable (e.g., in social
systems). For this reason, it is important to investigate how robust
the behaviour found under deterministic update rules is when noise is
added.  Examples for such studies are Ising-like models placed on a
network topology \cite{indekeu03,aleksiejuk02}, BN with a
probabilistic rule for choosing the update function at each time
step\cite{shmu02}, models with stochastic update sequences
\cite{greil05} and small stochastic delays in the update time
\cite{klemm05}. These stochastic models can lead to surprising new
results. For instance, for stochastic update sequences it could be
shown that the number of attractors (now defined as a recurrent set of
states) grows like a power law \cite{greil05,klemm05} as function of
the network size, while it grows superpolynomially for parallel
update.

In this paper, we investigate the effect of a perturbation on a BN
that is updated in parallel and that is on an attractor. The quantity
we evaluate is the probability that the network returns to the same
attractor after the perturbation, as a function of the size of the
perturbation, i.e., of the number of nodes the states of which are
changed. This leads to a curve that is characteristic of the network
and is strikingly different for different types of networks.

The outline of this paper is as follows: First, we investigate two
types of simple networks, namely networks consisting of independent
nodes and networks consisting of a single loop of nodes. The results
then help to understand the behaviour of Random Boolean Networks under
perturbations, which are studied in Section \ref{rbn}
dealing with frozen, critical and chaotic networks. In Section
\ref{specific}, we then investigate a few specific networks, which are not
random networks, and we find that their characteristic curves are very
different from those of random networks, reflecting for instance the higher
robustness to perturbations.  Finally, we summarize and discuss our
findings in Section \ref{conclusions}.

\section{Simple networks}

\subsection{Independent nodes}

Let us first consider a system of $N$ independent nodes. Then the
response of each node to the perturbation is independent of the
response of the other nodes. The state of each of the $N$ nodes at
time $t+1$ is determined by its state at time $t$. There are 4
possibilities to assign to such a node an update function: (i) the state
of the node is 0 irrespective of the state at the previous time step;
(ii) the state of the node is 1 irrespective of the state at the
previous time step; (iii) the state of the node at time $t+1$ is
identical to the state at time $t$; (iv) the state of the node at time
$t+1$ is the opposite of the state at time $t$. This means that the
node alternates between 0 and 1. 
The first two update functions are constant functions, the third is
the ``copy'' function, the fourth the ``invert'' function.

We generate a network by assigning update functions to the $N$ nodes. 
We then initialize the network in a randomly chosen state and wait until it
reaches an attractor. In the simple networks considered here, an attractor is
reached after one time step. Unless no ``invert'' function is
chosen, the period of the attractor is 2. When a node is perturbed,
its response depends on its update function: If the update function is
constant, the node returns in the next time step to its previous
value. If the update function is ``copy'', the node remains in its new
state and does not return to its previous state. If the update
function is ``invert'', the node continues to oscillate between 1 and 0,
however with a phase shift of one time step. 

When we perturb $h$ nodes, the network returns to the attractor only
if all perturbed nodes have a constant function, or if all nodes with
an ``invert'' function are perturbed but none with a ``copy''
function. Let us focus on the case that $N$ is large and that each of
the  four update functions is assigned to one quarter of the nodes. 
The probability that the network returns to its previous attractor
after perturbing $h$ nodes is then
\begin{equation}
P_{ret}(h) = \frac{{{N/2} \choose h}}{{N \choose h}} \, ,
\label{formelindep}
\end{equation}
which can be approximated by 
\begin{equation}
P_{ret}(h) \simeq \frac{1}{2^h}
\label{formelindepa}
\end{equation}
for $h/N \ll 1$. 
For $h > N/2$, the return probability is 0, since it is no longer
possible that all perturbed nodes have constant functions. For $N/4
\le h \le 3N/4$, it is in principle possible that all nodes with
``invert'' functions but none with ``copy'' functions are perturbed,
however, the probability that this occurs is so small that we neglect
it here. 

For general networks, the return probability has to be evaluated by
averaging over different initial states, however, in the special case
considered here the return probability is the same for each initial
state.

\begin{figure}[H]
\begin{center}
\includegraphics[width= \imgwidth \columnwidth ]{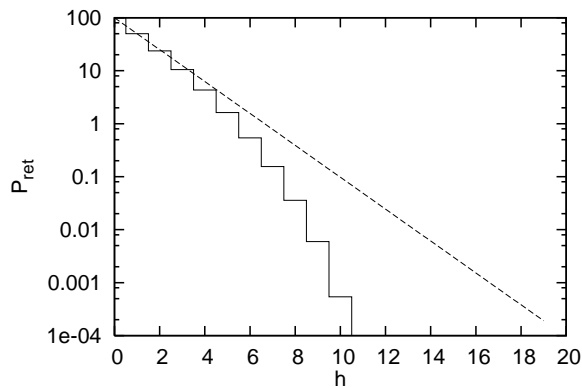}
\end{center}
      \caption{$P_{ret}(h) $ (in percent) of a set of independent nodes as given by Eq.~(\ref{formelindep}), the dashed line is obtained using Eq.~(\ref{formelindepa})}
	\label{indep}
\end{figure}

The most important result of this subsection is that for small
perturbations the return probability is simply given by an exponential
dependence
\begin{equation*}
P_{ret}(h)=P_{ret}(1)^h\, .
\end{equation*}
This result will be generally true for small $h$ whenever
perturbations at different nodes decay independently from each
other. For this reason, we will see an exponential decrease in
the examples below when $N$ is large and $h$ is small.

\subsection{Simple loops}

Now, let us consider a simple loop of $N$ nodes. 
\begin{figure}[H]
\begin{center}
      \includegraphics[width=0.5 \columnwidth]{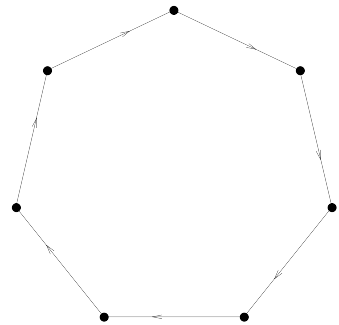}
\end{center}
      \caption{A loop of seven nodes}
	\label{loop}
\end{figure}
Each node is connected to its predecessor on the loop, and the update
function of each node is either ``copy'' or ``invert''.  A constant
function at one node will freeze the whole loop, in which case
$P_{ret}=1$ for all $h$, and we therefore focus on the more
interesting case of no constant function in the loop.  A loop with $n$
inversions can be mapped bijectively onto one with $(n-2)$ inversions
by replacing two ``invert'' with two ``copy'' functions and by
inverting the values of all nodes between these two couplings.  It is
therefore sufficient to distinguish loops with an even or an odd
number of inversions, and we call them ``even'' and ``odd'' loops
respectively. When discussing even loops, we consider loops with only
``copy'' functions. To odd loops we assign one ``invert'' function and
$N-1$ ``copy'' functions. 
An even loop with a prime number of
nodes returns to its initial state after $N$ time steps.  If $N$ is
not prime, shorter periods exist. Furthermore, if all nodes
have the same state, the loop is on a fixed point. 
An odd loop with a prime number of nodes returns to its initial state after $2N$
time
steps. An odd loop has no fixed points. The shortest
attractor has period 2, with alternating 1's and 0's.

\begin{figure}[H]    
\begin{center}
\includegraphics[width=\imgwidth \columnwidth]{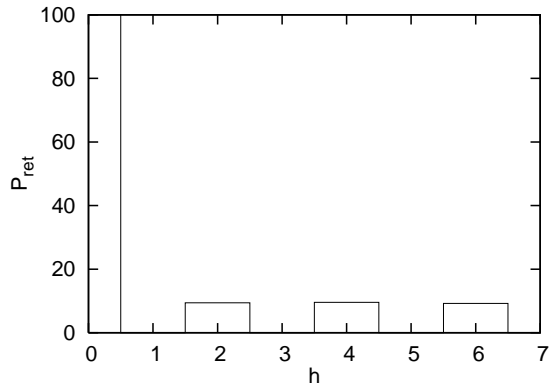}
\end{center}
      	\caption{$P_{ret}(h) $(in percent) for an even loop with $N=7$}
	\label{even}
\end{figure}

Figure \ref{even} shows $P_{ret}$ for an even loop with $N=7$, as
obtained from computer simulations.
 	
The return probability is 0 for all odd values of $h$, and it has the
same nonzero value for all even $h$ larger than 0. This happens for all
even loops with $N$ being a  prime number, as we show in the following. 
If $h$ nodes are perturbed, the loop returns to the same attractor
only if the perturbation leads it to another state of this attractor. 
All states of the same attractor have the same sequence of 0s and 1s, 
but rotated around the loop by some number of steps. Only by inverting
an identical number of 0s and 1s can we stay on the same
attractor, and this explains why $P_{ret}$ is nonzero only for even
$h$. Now, by fixing the nodes that are to be perturbed and by fixing
the number $m$ of steps by which the perturbation of these nodes shall
rotate the attractor, we uniquely fix the attractor (apart from an
inversion of all nodes). This attractor is found by fixing the state
of one node to 1 (or 0) and by requiring that the node $m$ steps ahead has
the same state if the first node is not a perturbed node, and the
opposite state if the first node is a perturbed node. Then we
determine in the same way the state of the node $m$ steps further,
and so on, until the state of all nodes is fixed. Since $N$ is a prime
number, we do not return to the first node before all other nodes have
been visited. The probability to be still on the same attractor after
perturbing $h$ nodes is for even $h$ therefore 
\begin{equation}
P_{ret} = \frac{2(N-1)}{2^N} \, ,
\end{equation}
which is the number of configurations that are
rotated by $m=1,\dots N-1$ steps under the perturbation, divided by
the total number of configurations of the loop.
If $N$ is not a prime number, there are attractors of different
length, and the function $P_{ret}(h)$ depends in a more complicated way on $N$. 

\begin{figure}[H]      
\begin{center}
\includegraphics[width=\imgwidth \columnwidth]{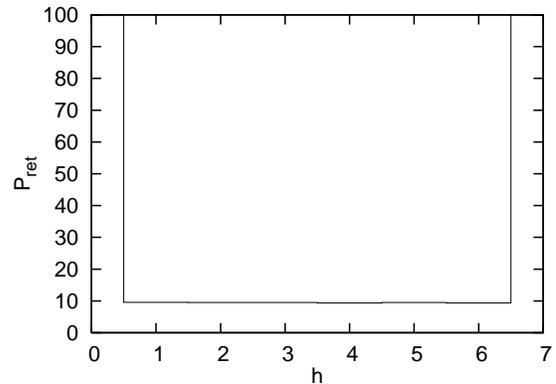}
\end{center}
      			\caption{$P_{ret}(h) $(in percent) of an odd loop $N=7$}
\end{figure}

For an odd loop with prime $N$, $P_{ret}$ is 1 for $h=0$ and $h=N$,
and has otherwise the value $P_{ret} = 2(N-1)/2^N$. This result is
obtained in a similar way as for even loops: The loop remains on the
same attractor after perturbing $h$ nodes, if the state generated by
the perturbation is among the $2N-2$ different states assumed by the
loop during the next $2N-1$ time steps, but not after $N$ steps (where
the state of all nodes is simply inverted, which can only be achieved
for $h=N$). If we fix the nodes to be perturbed and the number $m$ of
time steps between the unperturbed and the perturbed state, we can
uniquely identify the attractor by fixing the state of one node and
then stepping around the loop in steps of size $m$, assigning to each
node the same state as to the previous node if the previous node is
not among the perturbed nodes and if the ``invert'' function is not
between the two nodes. Otherwise, the inverted state is assigned.

\subsection{Collections of several loops}

If the network consists of several independent loops, the periods of
the attractors are the least common multiples of the periods of the
loops. Obviously, a collection of loops can remain on the same
attractor after a perturbation only if the perturbation produces a
state that corresponds to a future state of the loop, but not a state
that is not part of the state sequence on this loop.  If all loop
lengths are different from each other and have no common divisor,
every combination of future loop states belongs to the same attractor,
and the return probability is identical to the probability that the
perturbation generates only such states. Otherwise, if there are two
or more loops with periods that have a common divisor, the
perturbation must advance all these loops by a multiple of this
divisor if the system shall remain on the same attractor.  Since the
analytical expressions for $P_{ret}$ become complicated and do not
provide special insight, we omit them here. We only note that the
function $P_{ret}(h)$ depends strongly on the lengths and number of
loops, and on the common divisors of their lengths.

\section{Random Boolean Networks}
\label{rbn}

A random Boolean network (RBN) is constructed by choosing for each
node at random $k$ nodes from which it receives its input, and an
update function that assigns to each of the $2^k$ states of the $k$
input nodes an output 1 or 0. The update function of each node is
chosen at random among all $2^{2^k}$ possible update functions
\cite{kau1969a,kau1969b}.  All nodes are updated in
parallel. Depending on the value of $k$ and the probabilities assigned
to the different update functions, the dynamics of the network is
either in the frozen or in the chaotic phase. At the boundary between
the two are critical networks. If all update functions are assigned
the same probability, RBNs with $k=1$ are in the frozen phase,
networks with $k=2$ are critical, and networks with $k>2$ are chaotic.
 In the following, we consider
perturbations of these three types of networks.

\subsection{RBNs in the frozen phase}

In the frozen phase, all nodes apart from a small number (that remains
finite in the limit of infinite system size) assume a constant value
after a transient time.  If in the stationary state the value of one
node is changed, this perturbation propagates during one time step on
average to less than one other node.  If all nodes that become frozen
after some time are removed from the network, there remain the
nonfrozen nodes. In the limit $N \to \infty$, it can be shown that
these nonfrozen nodes are connected to simple loops (the ``relevant
loops'') with trees rooted in the loops \cite{kaufman05b}. The nodes
in the trees are slaved to the dynamics on the loops, and the
perturbation of a node on a tree does not induce a change of the
attractor. The perturbation of a frozen node can affect a couple of
other nodes to which the perturbation may propagate. If these other
nodes are also frozen or part of a nonfrozen tree, they will soon
return to the behaviour they showed before the perturbation. However,
if the perturbation affects a node sitting on a relevant loop, the
attractor will usually be changed. If we denote with $p_r$ the
probability that the perturbation of one node will affect a node on a
relevant loop, we obtain $P_{ret}(h) = (1-p_r)^h$ for small $h$. In
the limit $N\to\infty$, the probability $p_r$ must become proportional
to $1/N$, since the number of nodes on relevant loops remains finite
in this limit.

From these properties of frozen networks in the limit of large $N$, we
can conclude that $P_{ret}(h)$ is close to 1 for $h \ll N$, since the
probability of perturbing one of the relevant nodes is very
small. When $h$ becomes of the order of $Np_r$, relevant nodes are
perturbed with a considerable probability, and the shape of the
function $P_{ret}(h)$ depends on the properties of the relevant loops.
The size and number of relevant loops is in general different in
different networks. From our discussion in the previous section for
collection of independent loops, we conclude that there is no
self-averaging of $P_{ret}(h)$ in the limit of large $N$.  Also, the
probability that the number of relevant nodes is 0 approaches a
nonzero constant in the limit $N\to \infty$, and therefore some
networks always have a constant return probability $P_{ret}(h) = 1$.

\begin{figure}[H]
\begin{center}
     \includegraphics[width= \imgwidth \columnwidth]{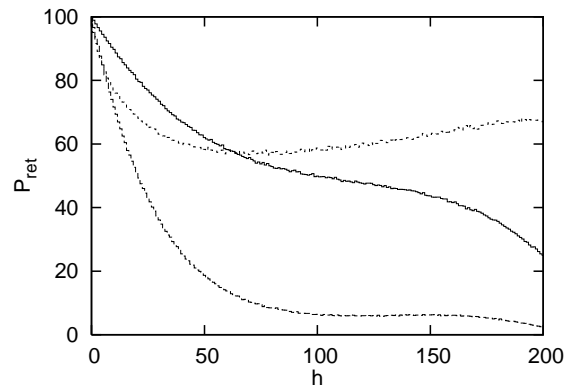}
\end{center}
      \caption{$P_{ret}(h) $(in percent) of three frozen networks $N=200, \langle k\rangle=1.5$}
	\label{frozen}
\end{figure}

Figure \ref{frozen} shows the result of a computer simulation of three
networks in the frozen phase. Half of the nodes were assigned the
value $k=1$ and half of the nodes the value $k=2$. (This combination
of $k$ values was chosen because networks with $k=1$ are almost always
completely frozen. Even for the value $\langle k \rangle = 1.5$, only
few networks were not completely frozen.)  One can see the inital
exponential decay and the broad range of curve shapes when $h/N$ is
not small anymore.  The behaviour of the curves when $h$ approaches
$N$ depends on the probability that the inversion of the state of all
nodes leaves the network in the same basin of attraction.  As we can
see in this and other figures in this paper, this probability varies
widely between networks and is not rarely larger than $P_{ret}(N/2)$.

\subsection{RBNs in the chaotic phase}

In the chaotic phase, initially similar configurations diverge
exponentially. Attractors are usually long, and a non-vanishing
proportion of all nodes keep changing their state even after long
times. Networks with $k=N$ are easiest to understand among the chaotic
networks, since the successor of each network state is a randomly
chosen other network state \cite{aldaco}. In state space, we have
therefore a ``network'' with one randomly chosen ``input''
(i.e. successor) for each ``node'' (i.e., state). In such a $k=N$
network, every perturbation leads the network anywhere in state space,
and we therefore expect $P_{ret}(h)$ to be a constant function, with
the exception of the value 1 at $h=0$. The average value of this
constant can be determined from what is known for $k=N$ networks: 
The value of $P_{ret}$ on the plateau is given by the average
probability that a randomly chosen state in state space belongs to the
same basin of attraction as the randomly chosen initial network state.
We define the weight $w_\rho$ of an attractor $\rho$ as the length of the
attractor plus the number of basin states draining into that attractor,
normalized by the size of the state space ($2^N$), so that $\sum_\rho w_\rho =
1$ \cite{basin}. We therefore have
\begin{equation}
P_{ret}=\sum\limits_{\rho} w_\rho^2
\label{formelbasins}
\end{equation} 

In the limit $N \to \infty$, the average number of attractors with a weight
between $w$ and $w+dw$ is given by \cite{derrida,basin}

\begin{equation}
g(w) = \frac{1}{2w\sqrt{1-w}} \, ,
\end{equation} 
leading to
\begin{equation}
P_{ret}= \int_0^1 g(w) w^2 dw = \frac 2 3 \, .
\label{formelbasinsinfty}
\end{equation} 
This analytical result applies to the ensemble average of all networks
with $k=N$. In a given network, the value of the plateau usually
deviates from 2/3, since always a few large basins dominate the value
of $P_{ret}$, and the values $w$ of these basins differ between
networks even in the thermodynamic limit.

\begin{figure}[H]
\begin{center}
\includegraphics[width=\imgwidth \columnwidth]{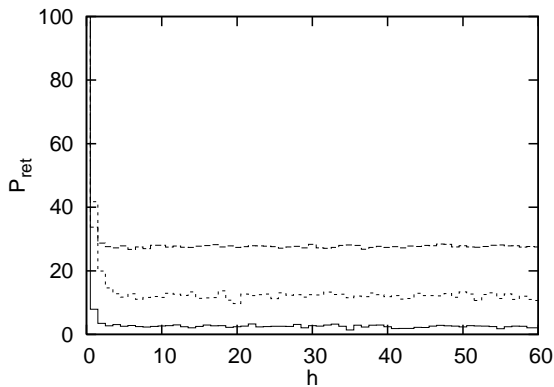}
\end{center}
	\caption{$P_{ret}(h) $(in percent) of three chaotic networks with $N=60,k=3$ }
	\label{chao}
\end{figure}

Now, let us turn to chaotic networks with fixed $k$ (smaller than
$N$), for which no such analytical results are known.  Figure
\ref{chao} shows $P_{ret}$ as function of $h$ for three networks with $k=3$
and $N=60$. We see a plateau for $h>2$, which means that perturbing
one or two nodes does not yet necessarily carry the network to a
random state, but perturbing several nodes does. The reason for this
is that a certain proportion of nodes in chaotic networks with fixed
$k$ assume a constant value after some time, and perturbing only such
nodes may not change the attractor. For larger $h$, it becomes very
likely that relevant nodes are changed by a perturbation, and then the
situation is similar as in $k=N$ networks, which have no frozen core.

\begin{figure}[H]
\begin{center}
     \includegraphics[width=\imgwidth \columnwidth]{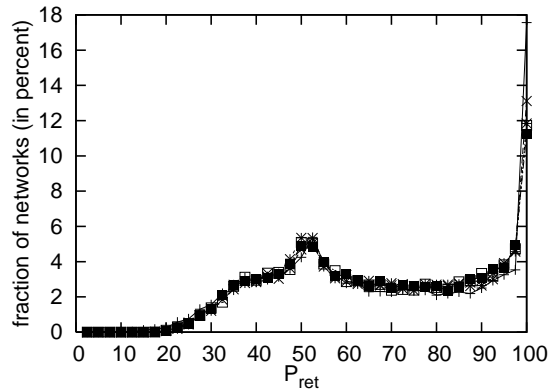}
\end{center}
      \caption{Probablity distribution of the value $P_{ret}(10)$ for approximately 10000 networks of size $N=20$, for different values of $k$.}
	\label{dist-pret}
\end{figure}

Chaotic networks with fixed value of $k$ and large $N$ have been shown
to share many properties with chaotic $k=N$ networks.  From
\cite{basin}, it appears that chaotic networks with $k< N$ have the
same set of basin weights as those with $k=N$, which would mean that
the average height of the plateau should be the same for all chaotic
networks in the thermodynamic limit.  We tested this assumption by
simulating smaller networks (of size 20) on the computer for $k=3$,
$4$ and $5$. The average of the return probability for a perturbation
of size $N/2 = 10$ over approximately 10000 networks is
$P_{ret}(10)=0.6764$, $0.675$, $0.6735$, $0.6686$ and $0.6675$ for $k=
3,4,5,6,9$ respectively.  These values are impressively close to 2/3
given the small system size.  For larger $k$, the value is closer to
$2/3$, because the number of nonfrozen nodes and therefore the size of
the (nontrivial part of the) state space increases with increasing
$k$.  We also simulated systems with larger values of $N$, however, it
is very hard to obtain good statistics for larger $N$ because the size
of the state space and the length of attractors increase exponentially
with $N$. From the data we have it appears that the value of the
plateau moves closer to 2/3 with increasing $N$.  Figure
\ref{dist-pret} shows the probability distributions of $P_{ret}(10)$,
from which the above-mentioned averages have been obtained. This
distribution is very broad and does not appear to become narrower for
larger $N$. There is no self-averaging of the plateau value of
$P_{ret}$. 

  In summary, we find that
chaotic networks with large $N$ have a return probability $P_{ret}(h)$
that decays rapidly for small $h$ and then reaches a plateau, the
ensemble average of which lies at $2/3$.

\subsection{Critical RBNs}

\begin{figure}[H]
\begin{center}
     \includegraphics[width=\imgwidth \columnwidth]{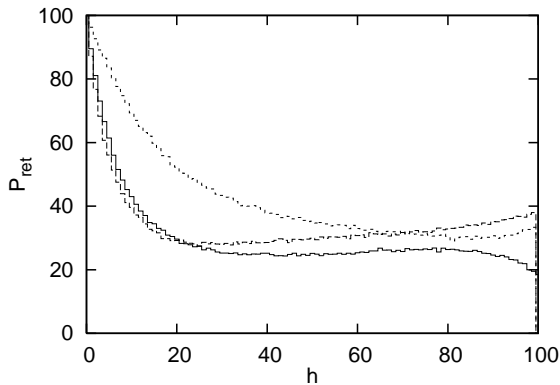}
\end{center}
      \caption{$P_{ret}(h) $(in percent) of three critical networks $N=100,k=2$}
	\label{krit}

\end{figure}

Critical networks are at the boundary between the frozen and chaotic
phase, and neighbouring configurations diverge only algebraically with
time.  Just as frozen networks, they have a large frozen core. For
large $N$, the number of nodes that do not become frozen after some
time, is proportional to $N^{2/3}$ \cite{kaufman05b}. Most of them are
irrelevant for determining the attractors. When the frozen core is
removed from the network, the remaining nodes are connected to
\emph{relevant components}, most of which are simple loops and all of
which contain loops, and with trees rooted in these loops.  The number
of nodes in the relevant components is proportional to $N^{1/3}$ for
large $N$. The mean number of nodes affected by the perturbation of a
single node is $\sim N^{1/3}$ for large $N$. The probability that a perturbation of size $h=1$ affects a relevant node is given by the probability $\sim N^{-2/3}$ that a given node is affected by the perturbation, times the number of relevant nodes, $P_{ret}(1) \simeq aN^{-1/3}$.
We therefore obtain
\begin{equation}
P_{ret}(h) \simeq
(aN^{-1/3})^h
\end{equation}
 for $h \ll N^{1/3}$. This exponential decay can be seen in Figure
\ref{krit} for small $h$.  For $h > N^{1/3}$, the probability of
perturbing a relevant node is not small anymore, and $P_{ret}$ reaches
a plateau. A rough estimate of the dependence on $N$ of the value of
$P_{ret}$ on the plateau is obtained by the following reasoning: The
number of attractors increases with $N$ roughly as $2^{N^{1/3}}$, and
the basins of attraction have a size of the order $2^N/2^{N^{1/3}}$.
Using Eq.~(\ref{formelbasins}), this gives the estimate $P_{ret} \sim
2^{-bN^{1/3}}$ on the plateau with some constant $b$. The height of
the plateau decreases rapidly with increasing $N$ because of the vast
number of different attractors of comparable basin size.

Taking these results together, we expect that in large critical
networks the function $P_{ret}(h)$ decays from 1 to a value close to 0
when $h$ increases from 0 to $N^{1/3}$ and then stays on this plateau
as $h$ increases further. The value of plateau decreases as an
exponential function of $-N^{1/3}$.  However, it is known that the
scaling with $N^{2/3}$ and $N^{1/3}$ of the nonfrozen and relevant
nodes becomes clearly visible only in huge networks (of the order
$10^6$), and smaller networks may show broad distributions in the
numbers of these nodes. Figure \ref{krit} shows three examples of
$P_{ret}$ for critical networks. For such small networks sizes, the
critical features of $P_{ret}(h)$ derived for the thermodynamic limit
cannot yet be seen.

\section{Some specific networks}
\label{specific}

In the ensemble of all random networks of a not too large size, there are some
networks whose characteristic curves are very different from
conventional ones, as shown in the next two figures.

\begin{figure}[H]   
\begin{center}
	\includegraphics[width=\imgwidth \columnwidth]{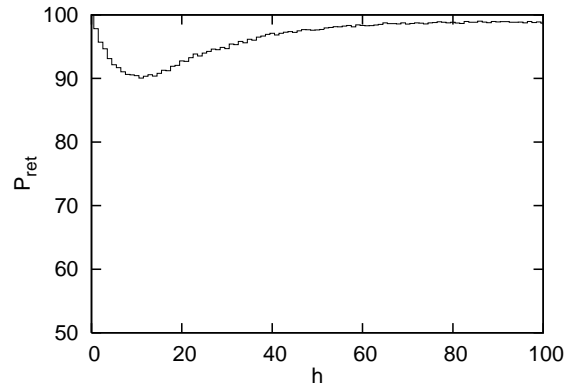}
\end{center}
      	\caption{$P_{ret}(h) $(in percent) of a critical network with $N=100,k=2$}
\end{figure}

\begin{figure}[H]      
\begin{center}
	\includegraphics[width=\imgwidth \columnwidth]{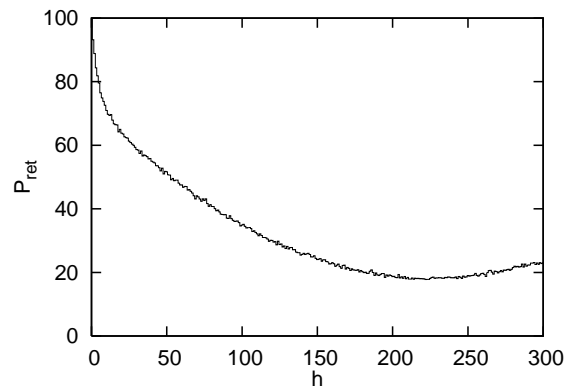}
\end{center}
	\caption{$P_{ret}(h) $(in percent) of another critical network with $N=300,k=2$}
\end{figure}
 	
The first curve is that of a very robust network, which additionally
has the property of returning more easily to the original attractor
when the perturbation is larger. 

The second curve is that of a network that has no plateau but a very
extended decrease of $P_{ret}$ with $h$. A closer inspection of this
network reveals that, although its parameter values classify it as
critical, there is no frozen core but all nodes are relevant and part
of a single complex relevant component. This means that the state
space is far from random, and that perturbations of different sizes
carry the network to different regions in state space. 

\begin{figure}[H]      
	
\begin{center}
\includegraphics[width=\imgwidth \columnwidth]{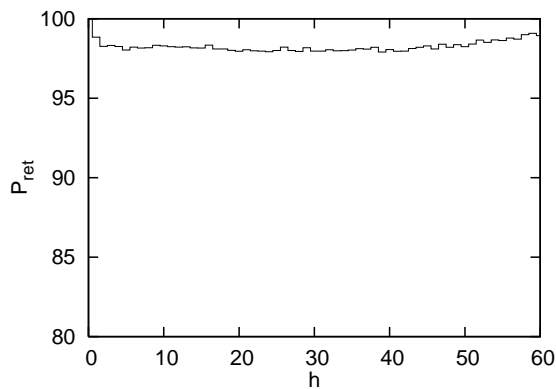}
\end{center}
      			\caption{$P_{ret}(h) $(in percent) of a network of 60 nodes evolved for robustness.}
			\label{agnes}
\end{figure}

Let us compare the first curve to that of networks that have been
evolved for robustness to small perturbations \cite{agnespaper}, shown in Figure
\ref{agnes}. These networks have been evolved by modifying the
connections and functions and by selecting for a high probability of
returning to the same attractor after perturbing one
node. Interestingly, although only a large value of $P_{ret}(1)$ has
been imposed in the system, $P_{ret}$ is very high for all
perturbation sizes. Many networks obtained
by the same procedure even show $P_{ret}(h)=1$ for all $h$.

\begin{figure}[H]      
\begin{center}
\includegraphics[width=\imgwidth \columnwidth]{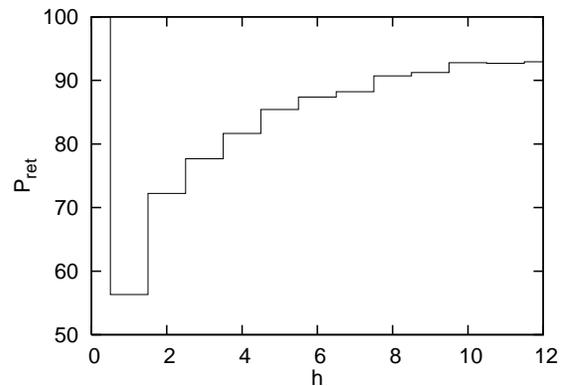}
\end{center}
      		\caption{$P_{ret}(h) $(in percent) of the cell cycle regulation network of budding yeast}
		\label{yeast}
\end{figure}

The network used for producing Figure \ref{yeast} is the core part of
cell cycle regulation network of budding yeast, as represented in
\cite{yeast}. An important property of this network is that it has a
prominent fixed point attracting $86\%$ of all network states. The
diagram shows the surprising feature that $P_{ret}(h)$ has its minimum
at $h=1$.  This means that the main attractor, which is a fixed point,
is not very stable under perturbations of one node. Further analysis
shows that there are three other fixed points that differ from the main fixed point by the state of only one node. As suggested in \cite{yeast}, 
the cell is likely to be waiting for a new input when the network is at its fixed point. This might explain why this fixed point is rather sensitive to small
perturbations.

\section{Conclusions}
\label{conclusions}

In this paper, we have evaluated the probability $P_{ret}(h)$ that a
Boolean network returns to an attractor after perturbing $h$ nodes,
averaged over different initial states of the network. We found that
$P_{ret}(h)$ can display a variety of different shapes, which yields
insights into the state-space structure. If the response of the
network to the perturbation of several nodes is independent for each
perturbed node, $P_{ret}(h)$ decays exponentially with $h$ for small
$h$. Larger perturbations are of course not independent, and if the
perturbation leads the system to a random place in state space,
$P_{ret}(h)$ shows a plateau for these perturbation sizes. When the
size of the perturbation approaches the total number of nodes,
$P_{ret}(h)$ can shown various types of behaviour, including a
decrease to zero or an increase to a value larger than the plateau.
The reason is that the response to inverting the state of all nodes
depends strongly on the network structure. We obtained analytical
results for Random Boolean Networks in the limit of large $N$. For
critical networks, $P_{ret}(h)$ decreases rapidly to (almost) 0 for
large $N$.  For chaotic and frozen networks, $P_{ret}(h)$ remains
nonzero for large $N$ and is not self-averaging, which means that the
shape of $P_{ret}(h)$ differs widely between different networks.  The
ensemble average of the plateau value of $P_{ret}$ for chaotic
networks is 2/3 for large $N$. The fact that this plateau value is
much higher than in critical networks means that chaotic networks are
more robust than critical networks when perturbations affect a
nonvanishing proportion of all nodes. The reason for this is that
chaotic networks have a much smaller number of attractors than
critical networks, and therefore the probability that a perturbation
carries the network into the basin of attraction of a different
attractor is larger in critical networks than in chaotic networks.
Similarly interesting and surprising is the result that a biological
network that has been characterized as being very robust is pretty
sensitive to perturbations of size $h=1$, while $P_{ret}$ is much
larger for larger perturbations. It remains to be seen if this feature
occurs also in other biological networks.

To conclude, characterizing the dynamical stability of networks by a function
$P_{ret}(h)$ gives far more information about the state space
structure of the network than using a single number, such as the
average ``sensitivity'' of nodes.

\section{Acknowledgments}
We thank C. Hamer for programming the yeast network and 
T. Mihaljev for useful discussions.

\end{document}